\documentclass[12pt,fleqn]{article}
\usepackage{amsmath,a4,latexsym,epsfig}
\newcommand{\intd}{\int \! d^4 x \;}
\newcommand{\intS}{\int \! d S \;}
\newcommand{\intSbar}{\int \! d \bar S \;}

\newcommand{\Ga} {\Gamma}
\newcommand{\Gacl} {{\Gamma_{\rm cl}}}

\renewcommand{\L}{{\cal L}}

\newcommand{\lambdabar}{{\overline\lambda}}

\newcommand{\psibar}{{\overline\psi}}

\newcommand{\epsilonbar}{{\overline\epsilon}}
\newcommand{\thetabar}{{\overline\theta}}
\newcommand{\etabar}{{\overline\eta}}
\newcommand{\chibar}{{\overline\chi}}

\newcommand{\fbar}{{\overline f}}
\newcommand{\cbar}{{\overline c}}

\newcommand{\alphadot}{{\dot\alpha}}
\newcommand{\betadot}{{\dot\beta}}

\newcommand{\Tr}{{\rm Tr}}

\def\df#1{\frac{\delta}{\delta#1}}

\def\pslash#1{{\setbox0=\hbox{$#1$}
  \rlap{\ifdim\wd0>.7em\kern.22\wd0\else\kern.1\wd0\fi /}#1}}

\def\brs{\mathbf s}
\newcommand{\etabarbold}{{\mbox{\boldmath{$\etabar$}}}}
\newcommand{\etabold}{{\mbox{\boldmath{$\eta$}}}}
\newcommand{\mn}{{\mu\nu}}
\newcommand{\rs}{{\rho\sigma}}

\begin{document}
\begin{titlepage}

\begin{flushright}
October 2002\\
BN--TH--05--2002\\
hep-th/0211084
\end{flushright}
\vspace{8ex}
\begin{center}
{\large\bf{
        Anomalies in quantum field theory\\[1ex]
         Properties and characterization}}
\\
\vspace{8ex}
{\large       E. Kraus} 
{\renewcommand{\thefootnote}{\fnsymbol{footnote}}
\footnote{Talk presented at the Hesselberg workshop 2002 ``Renormalization
and regularization'',\\ 22 Feb. to 1 March 2002.
}} 
\\
\vspace{2ex}
{\small\em                Physikalisches Institut,
              Universit{\"a}t Bonn,\\
              Nu{\ss}allee 12, D--53115 Bonn, Germany \\
              E-mail: kraus@th.physik.uni-bonn.de\\}
\vspace{2ex}
\end{center}
\vfill
{\small
 {\bf Abstract}
 \newline\newline
We consider the Adler-Bardeen anomaly of the  $U(1)$ axial current in
abelian and non-abelian gauge theories and present its algebraic
characterization as well as an explicit evaluation proving 
regularization scheme independence of the anomaly. By extending the
gauge coupling to an external space-time dependent field we get a
unique definition for the quantum corrections of
 the topological term. It also implies a simple proof of the 
non-renormalization theorem of the Adler-Bardeen anomaly.
We consider local gauge couplings in supersymmetric theories
and find that there the renormalization of the gauge coupling is determined
by the topological term in all loop orders  except for one loop. It is
shown that in one-loop order the quantum corrections to the
topological term induce an anomalous breaking of supersymmetry, which
is characterized by similar properties as the Adler-Bardeen anomaly.

\hfill
}

\end{titlepage}


\section{Introduction}

Anomalies appear as a breaking of classical symmetries in quantum
field theory. They arise from divergent loop diagrams and they
 cannot be removed by adjusting counterterms to the classical
action. As such they are seen to be true quantum effects appearing in
the procedure of renormalization and quantization of the classical
field theory.

In the first part of the  paper we consider the Adler-Bardeen anomaly
\cite{AD69,BA69,BEJA69}
of softly broken
$U(1)$ axial symmetry in abelian and non-abelian gauge theories.
The Adler--Bardeen anomaly is determined by the topological term
$\Tr\, G \tilde G$ and algebraically characterized as a
non-variation under gauge and BRS symmetry, respectively. We prove
explicitly that the coefficient of the anomaly is determined by
convergent one-loop integrals, which are not subject of
renormalization. As such its coefficient is independent from the
regularization scheme used for the subtraction of UV divergences.
Another characteristic property of the Adler-Bardeen anomaly is its
non-renormalization theorem
\cite{ADBA69}, which states, that the coefficient of the
anomaly is not renormalized in higher orders. We give a new and simple
proof of the non-renormalization theorem by extending the gauge
coupling to an external field. Then
 the renormalization of the
topological term is  unambiguously
determined  by gauge invariance and implies the non-renormalization theorem.  

In the second part we turn to supersymmetric gauge theories. It was
recently shown that with 
local coupling supersymmetric Yang-Mills theories have an anomalous
breaking of supersymmetry in one-loop order \cite{KR01}.
 Contrary to the Adler-Bardeen anomaly
the supersymmetry anomaly is a variation under BRS-symmetry, but the
respective counterterm depends on the logarithm of the coupling. Thus
its coefficient is scheme independent and is not introduced in the
procedure of renormalization. 

Evaluating the symmetry identities one can show, that the anomaly of
supersymmetry is induced by the renormalization of the topological
term \cite{KR01anom}. In addition we demonstrate that in all orders except for one
loop the renormalization of the topological term determines the
renormalization of the gauge coupling constant. As an application we
derive the closed expression of the gauge $\beta$ function in terms of
its one-loop coefficient and the anomaly coefficient.

\section{ The Adler-Bardeen anomaly} 

We start the considerations with the axial anomaly appearing in QED in
the axial current Ward identity
\cite{AD69}. The classical action of QED with one
Dirac spinor is given by:
\begin{eqnarray}
\label{Gacl}
\Gacl & = & -\frac 14 \intd F^{\mn} F_{\mn} - \intd \frac 1{2\xi}
(\partial^\mu A_\mu)^2 \nonumber \\
& & { }+ \intd \big\{ (i \psi_L \sigma ^\mu D_\mu \psibar_L + i 
 \psi_R \sigma ^\mu D_\mu \psibar_R ) - m (\psi_L \psi_R + \psibar_L
\psibar_R) \big\} 
\end{eqnarray}
with
\begin{eqnarray}
F^{\mn} & = & \partial^\mu A^\nu - \partial^\nu A^\mu \ ,\nonumber
\\
D_\mu \psi_P &=  &(\partial _\mu + i e Q_P A_\mu) \psi_P, \quad
P = L,R, \quad \mbox{and}  \quad Q_L = -1 , Q_R = +1\ , \nonumber \\
D_\mu \psibar_P &\equiv & (D_\mu \psi_P)^ \dagger 
\end{eqnarray}
Here, $\psi_L^\alpha $ and $\psi_R^\alpha $ are two-component Weyl
spinors and $\psibar_L^\alphadot $ and $\psibar_R^\alphadot $ are
their complex conjugates
\begin{equation}
\bigl( \psi_L^\alpha \bigr)^\dagger = \psibar_L^\alphadot\, .
\end{equation}
The left- and right-handed Weyl spinors compose the 4-component Dirac
spinor
according to
\begin{equation}
\Psi = \left( \begin{array}{c}  \psi_{L\alpha} \\ \psibar_R^\alphadot
\end{array}\right) \ .
\end{equation}
Up to the gauge fixing term 
$(\partial^\mu A_\mu)^2 $ the classical action (\ref{Gacl})
is invariant under
$U(1)$ gauge transformations:
\begin{eqnarray}
\label{gaugetrafo}
\delta^{\rm gauge} _\omega A_\mu & = &\frac 1e \partial_\mu \omega \ ,\nonumber
\\
\delta^{\rm gauge} _\omega \psi_P & = & -i \omega Q_P \psi_P \ .
\end{eqnarray}
$U(1)$ gauge invariance is expressed by the gauge Ward identity
\begin{equation}
\label{WIgauge}
\Bigl( {\bf w}^{\rm gauge} - 
 \partial^\nu \frac 1e \df{A^\nu} \Bigr) \Gacl = 
- \frac 1 {\xi e} \Box \partial A\ ,
\end{equation}
with the Ward operator
\begin{equation}
{\bf w}^{\rm gauge} \equiv \sum_{P = L,R}
\Big(\delta^{\rm gauge} \psi^\alpha _P
\df{\psi^\alpha_P} + \delta^{\rm gauge} \psibar^\alphadot _P
\df{\psibar^\alphadot_P} \Big)\ .
\end{equation}

QED is renormalizable which means that
 the Ward identity (\ref{WIgauge}) can be
fulfilled to all orders in perturbation theory in its classical form:
\begin{equation}
\label{gaugeWI}
\Bigl( {\bf w}^{\rm gauge} -  \partial^\nu \frac 1e \df{A^\nu} \Bigr) \Ga = 
-\frac 1 {e\xi} \Box \partial A\ ,
\end{equation}
where $\Ga = \Gacl + O(\hbar)$ is the generating functional of
   one-particle-irreducible (1PI) Green functions.
The Ward identity together with a subtraction procedure for removing
the ultraviolet divergences defines the 1PI Green functions of QED.

Having specified the vertex function of QED one can  consider
the action of axial transformations on the 1PI Green functions of QED.
Axial symmetry is classically broken by the mass term of matter
fields and its local version implies partial conservation of the axial current.
We introduce in the classical action
(\ref{Gacl}) an external vector field $V^\mu$
coupling to the axial current, i.e.\  we
extend the covariant derivatives in (\ref{Gacl}) to,
\begin{equation}
D^\mu \psi_P = (\partial^\mu + i e Q_P A^\mu + i V^\mu)\psi_P \ .
\end{equation}
Then softly broken axial symmetry can
be expressed in form of a softly broken axial Ward identity
\begin{equation}
\label{WIaxial}
\Bigl( {\bf w}^{\rm axial} - \partial^\nu \df{V^\nu} \Bigr) \Gacl = 
 2 i m (\psi_L \psi_R + \psibar_L \psibar_ R)
\end{equation}
with the Ward operator of axial transformations
\begin{equation}
\label{WOaxial}
{\bf w}^{\rm axial} \equiv  \sum_{P = L,R} \Bigl(-i  \psi^\alpha _P
\df{\psi^\alpha_P} + i  \psibar^\alphadot _P
\df{\psibar^\alphadot_P} \Bigr)
\end{equation}

In one-loop order the axial Ward identity is broken by the
Adler-Bardeen anomaly:
\begin{equation}
\Bigl( {\bf w}^{\rm axial} - \partial^\nu \df{V^\nu} \Bigr) \Ga^{(1)} = 
e^2 r^{(1)} \epsilon ^{\mn \rho \sigma } [F_{\mn} F_{\rho\sigma}]\cdot \Ga 
+ \mbox{soft terms} \ .
\end{equation}

The axial anomaly is characterized by the following properties:
\begin{enumerate}
\item
If the gauge Ward identity is established, then $\epsilon^{\mn\rho
\sigma} F_{\mn} F_{\rho \sigma }$ cannot be expressed as the variation of
a 4-dimensional field monomial. Indeed,  the only possible counterterm with the
$\epsilon$-tensor
\begin{equation}
\Ga_{\rm ct, noninv} = 4 r^{(1)}e^2 \intd \epsilon^{\mn\rho
\sigma} V_\mu A_\nu \partial_\rho A_\sigma
\end{equation}
is not gauge invariant. Adding it to the classical action the
anomaly is shifted to the gauge Ward identity, i.e.,
\begin{eqnarray}
\Bigl( {\bf w}^{\rm gauge} - \frac 1e\partial^\nu \df{A^\nu} \Bigr)
\Ga^{(1)}  & = &
r^{(1)} e \epsilon^{\mn\rho
\sigma} F_{\mn}(A) F_{\rho \sigma}( V)\ ,  \nonumber \\
\Bigl( {\bf w}^{\rm axial} - \partial^\nu \df{V^\nu} \Bigr) \Ga^{(1)} & = &
\mbox{soft terms}\ .
\end{eqnarray}
From its algebraic characterization one can immediately derive that
the coefficient of the anomaly $r^{(1)} $ is scheme-independent and independent
of the specific subtraction procedure used for renormalization of 1PI
Green functions.
\item
The anomaly is not renormalized \cite{ADBA69},
i.e.,
one has
\begin{equation}
\Bigl( {\bf w}^{\rm axial} - \partial^\nu \df{V^\nu} \Bigr) \Ga = 
e^2 r^{(1)} \epsilon ^{\mn \rho \sigma } [F_{\mn} F_{\rho\sigma}]_4
\cdot \Ga + 
 2 i m [\psi_L \psi_R + \psibar_L \psibar R]_3 \cdot \Ga\ .
\end{equation}
It is obvious that for a precise definition of the non-renormalization
theorem one first has to define  the Green functions with anomaly
insertions $ [F_{\mn} \tilde F^{\mn}]_4
\cdot \Ga$. A well-suited procedure for this definition is the usage of
local gauge couplings, which makes possible to define the
renormalization
of $\epsilon^{\mn\rs} F_\mn  F_{\rs} $
unambiguously by gauge-invariance
and to state the non-renormalization of the Adler-Bardeen anomaly
 in a scheme-independent framework.
\end{enumerate}
In the following subsections we will have a closer look on the
properties of the Adler-Bardeen anomaly.

\subsection{Scheme independence of the anomaly}

The algebraic characterization of the Adler-Bardeen anomaly implies,
that it is induced by convergent one-loop integrals, which are not
subject of renormalization \cite{AD69}. (For a recent review see
\cite{JE00}.)
 To work this property out more explicitely
we test the axial Ward-identity with respect to two photon fields and
obtain after Fourier transformation:
\begin{equation}
\label{waxialAA}
- i q^\rho \Ga_{V^\rho A^\mu A^\nu} (q,p_1,p_2) =  r^{(1)} 8
\epsilon^{\mn\rho \sigma} p_{1\rho} p_{2\sigma} + \mbox{soft terms.}
\end{equation}
Gauge invariance implies:
\begin{equation}
p_1^\mu \Ga_{V^\rho A^\mu A^\nu} (q,p_1,p_2) = 
p_2^\nu \Ga_{V^\rho A^\mu A^\nu} (q,p_1,p_2) = 0
\end{equation}
We carry out a tensor decomposition of the Green function 
$ \Ga_{V^\rho A^\mu A^\nu} (q,p_1,p_2)$ and find:
\begin{eqnarray}
\label{tensoraxial}
 \Ga_{V^\rho A^\mu A^\nu} (q,p_1,p_2)
& = & i \epsilon^{\rho \mu \nu \lambda} p_{1\lambda} \Sigma_{\rm
div}(p_1,p_2) 
- i \epsilon^{\rho \mu \lambda \lambda'}p_{2 \lambda} p_{1\lambda'} \sum_{i=1,2}
p^{\nu}_i 
\Sigma_i (p_1, p_2)\nonumber \\
& & { }+ (p_1 \leftrightarrow p_2, \mu \leftrightarrow \nu )  \ .
\end{eqnarray}
According to their momentum structure the scalar function $\Sigma_i(p_1,p_2)$
arise from convergent one-loop integrals. Using now transversality in
the photon legs we find that the divergent part is uniquely determined
by the convergent one-loop integrals:
\begin{equation}
\Sigma_{\rm div} (p_1,p_2) =  p_2^2 
\Sigma_2 (p_1,p_2) + p_1 \cdot p_2 \Sigma_1 (p_1,p_2)\ .
\end{equation}
Thus, in the end one finds that the anomaly coefficient is determined
in terms of the convergent one-loop functions $\Sigma_i$. Evaluating
(\ref{waxialAA}) for asymptotic moment much larger than the fermion mass, where the soft mass contribution vanishes, one obtains:
\begin{eqnarray}
8 r^{(1)} & = & \Sigma^{(1)}_{\rm div}(p_1,p_2 ) + \Sigma^{(1)}_{\rm
div}(p_2,p_1 )  \\
 & = & ( p^2_2 
\Sigma^{(1)}_2(p_1,p_2) + p_1^2\Sigma^{(1)}_2(p_2,p_1) \bigr)
 + p_1 \cdot p_2 \bigl( \Sigma_1(p_1,p_2)+
\Sigma_1(p_2,p_1) \bigr)\ , \nonumber 
\end{eqnarray}
which demonstrates  scheme and regularization independence of
$r^{(1)}.$ The explicit evaluation of integrals can be found in the
original paper~\cite{AD69}.

\subsection{The non-renormalization theorem and the local gauge coupling}

For proving the non-renormalization theorem of the Adler-Bardeen
anomaly \cite{ADBA69}
we define in a first step the Green functions $\epsilon^{\mn \rs}
[F_{\mn} 
F_{\rs}]\cdot \Ga$. For this purpose it is useful to extend the
coupling constant to an external space-time dependent field, the local
coupling $e(x)$ \cite{KRST01}. With local gauge coupling the gauge transformation of
the photon field (\ref{gaugetrafo}) is generalized to
\begin{equation}
\delta_\omega^{\rm gauge} A_\mu = \frac 1{e(x)} \partial_\mu \omega\ .
\end{equation}
The kinetic action of the photon field
\begin{equation}
\Ga_{\rm cl,kin} = \intd \Bigl( -\frac 1{4e^2(x)}F^{\mn}(eA) F_{\mn}(eA) \Bigr)
\end{equation}
is gauge invariant under transformations with local coupling and is
the unique extension of the ordinary gauge invariant action (\ref{Gacl}).
Renormalization of QED with local gauge coupling is performed in the same
way as with constant coupling. We treat the local coupling as an
external field and perform the limit to constant coupling for all
explicit diagrams. In particular, the perturbative expansion is a
power series in the coupling and the 1PI Green function satisfy the
topological formula in loop order $l$:
\begin{equation}
\label{topfor}
N_e \Ga^ {(l)} = (N_A + N_\psi + 2(l-1))\Ga^{(l)}\ .
\end{equation}
$N_e, N_A, N_\psi$ are the usual leg counting differential operators, e.g.
\begin{equation}
N_e = \intd e(x) \df{e(x)}\ ,
\end{equation}
which include for fermion fields the complex conjugate and a sum over
left and right-handed fields. 

In the following we will show that due to the extension to local gauge coupling
 the insertions
$[F^\mn \tilde F_\mn]$ into photon self energies are uniquely 
defined by gauge invariance
with local gauge coupling.  For this purpose it is
 useful to couple the anomaly to an
 external field
$\Theta (x)$  in the classical action (\ref{Gacl}), i.e.,
\begin{equation}
\label{Thetadef}
\Gacl \to \Gacl - \frac 14 \intd \Theta(x) F^\mn(eA) \tilde F_\mn(eA)\ , \qquad
\tilde F^\mn = \epsilon^{\mn\rs} F_{\rs}\ .
\end{equation}
Differentiation with respect to $\Theta$ yields the  Green
functions with insertions $F\tilde F$:
\begin{equation}
\df{\Theta(x)} \Ga = - \frac 14 [F^\rs(eA) \tilde F_\rs(eA)]\cdot \Ga \ .
\end{equation}
An additional characterization for the renormalization is provided by
 the property, that $F \tilde F$ couples is a total derivative and
 vanishes upon integration. Thus one has
\begin{equation}
\label{Thetader}
\intd \df{\Theta(x)} \Gacl = - \frac 14\intd F^\mn(eA) \tilde F_\mn(eA)
= 0 \ .
\end{equation}
This property can be trivially maintained in the procedure of
renormalization, i.e.,
\begin{equation}
\label{FtildeFint}
\intd \df{\Theta(x)} \Ga = - \frac 14 \intd [F^\rs(eA) \tilde F_\rs(eA)]\cdot
\Ga
= 0\ ,
\end{equation}
and  implies that the UV divergences corresponding to the
respective diagrams are  total derivatives. 

Considering now the possible invariant counterterms to the field
$\Theta$ in the classical action, we find as the only gauge invariant
terms satisfying the constraint (\ref{FtildeFint})
\begin{equation}
\label{GactTheta} 
\Ga^{(l)}_ {{\rm ct}, \Theta} = \intd z_{V\Theta}^{(l)} 
 \Theta(x) \partial^\mu ( e^{2l} j_\mu^{\rm axial})\ .
\end{equation}
It corresponds to a field redefinition of the  external axial
vector:
\begin{equation}
\label{Vdef}
V^\mu \to V^\mu -
 z_{V\Theta}^{(l)}e^{2l}\partial^\mu
 \Theta \ .
\end{equation}
Higher-order counterterms to $F \tilde F$ itself
vanish, since they
have the general form
\begin{equation}
\label{FtildeFct}
\intd \Theta(x) e^{2l}(x) F^\mn {(eA)}\tilde F_\mn {(eA)} \ ,
\end{equation}
and violate the property (\ref{FtildeFint}) for local gauge coupling. 
Hence, higher-order corrections to
$\Ga_{\Theta A^\mu A^\nu}$ are non-local
and
  are determined by gauge invariance and by the identity
(\ref{FtildeFint}).

For the explicit construction
 we consider  the Green functions 
once differentiated with respect to $e(x)$:
\begin{equation}
\lim_{e \to {\rm const.}}
\df{\Theta(z)} \df{ e(y)}
\df{ A^\mu(x_1)} \df{A^\nu (x_2)}
\Ga\Big|_{A,\psi= 0}\equiv \Ga_{\Theta e A^\mu A^\nu} (z, y, x_1,x_2)
 \ .
\end{equation}
 The gauge Ward identity (\ref{gaugeWI}) with local coupling
yields the identities $(q + p+ p_1 +p_2 = 0)$:
\begin{eqnarray}
\label{gaugeAA}
& & p_1^\mu\Ga_{\Theta  A^\mu A^\nu}({q,p,p_1,p_2}) 
= 0 \ , \nonumber\\ 
& & e  p_1^\mu \Ga_{\Theta e A^\mu A^\nu}({q,p,p_1,p_2}) -
p_1^\mu \Ga_{\Theta  A^\mu A^\nu}({q,p_1 + p,p_2}) = 0 \ .
\end{eqnarray}
Due to the construction the Green functions once differentiated with
respect to the local coupling are related to the Green functions with
constant coupling. From the topological formula (\ref{topfor})
one gets
\begin{equation}
\label{topforAA}
2(l +1)  \Ga^{(l)}_{\Theta A^\mu A\nu}(q,p_1,p_2) =
e \Ga^{(l)}_{\Theta e A^\mu A^\nu}({q,0,p_1,p_2}) \ .
\end{equation}
Using that Green functions with $F\tilde F$ insertions vanish for
vanishing incoming momenta at the insertion (see (\ref{FtildeFint}))
 we find the following tensor decomposition for general $ \Ga_{\Theta
e A^\mu A^\nu}$: 
\begin{eqnarray}
\label{tensorAA}
& & \Ga_{\Theta e A^\mu A^\nu}({q,p,p_1,p_2})   \nonumber \\
&  & = \epsilon^{\mn\rs} q_\rho p_{1 \sigma} \Sigma_{\rm div}(p_1,p_2,p)
\nonumber \\
& & { }\; 
+ \epsilon^{\mu \lambda \rs} q_\lambda p_{1 \rho} p_{2 \sigma} \big(
p_1^{\nu} \Sigma_1 (p_1,p_2,p) + p_2^{\nu} \Sigma_2 (p_1,p_2,p)
 + p^{\nu} \Sigma_3 (p_1,p_2,p)\big) \nonumber \\
& & { } \; + (\mu \leftrightarrow \nu, p_1 \leftrightarrow p_2)\, .
\end{eqnarray}
 From (\ref{topforAA}) we obtain for $\Ga_{\Theta AA}$ at constant coupling
\begin{eqnarray}
\label{GaThetaAA}
  \Ga^{(l)}_{\Theta A^\mu A^\nu}(q,p_1,p_2) &\equiv & 
\epsilon^{\mn\rs} p_{1 \rho} p_{2 \sigma} \Sigma^{(l)}_{\Theta}(p_1,p_2)
\nonumber \\
& = & \frac{e}{2(l+1)} \epsilon^{\mn\rs} p_{1 \rho} p_{2 \sigma}\big( 
\Sigma^{(l)}_{\rm div}(p_1,p_2,0) + \Sigma^{(l)}_{\rm div}(p_2,p_1,0) \big) \ .
\end{eqnarray}

 Applying the gauge Ward identity (\ref{gaugeAA})
to (\ref{tensorAA}) the local part 
$\Sigma_{\rm div}$ is determined by the convergent tensor integrals
for all loop orders $l \geq 1$ and according to 
(\ref{GaThetaAA}) one has at the same time determined  $\Ga_{\Theta AA}$:
\begin{eqnarray}
\label{GaThetaresult}
\Sigma^{(l)}_\Theta (p_1,p_2) &= & \frac e {2 (l+1)}
\big(\Sigma^{(l)}_{\rm div}(p_1,p_2,0) + 
\Sigma^{(l)}_{\rm div}(p_2,p_1,0)\big)  \nonumber \\
& = & - \frac e {2l} \Bigl(p_1\cdot p_2 \big(\Sigma^{(l)}_{1}(p_1,p_2,0) + 
\Sigma^{(l)}_{1}(p_2,p_1,0)\big) \nonumber \\
& & { }
\phantom{ - \frac 1{2l}}\;
   +p_2^2 \Sigma^{(l)}_{2}(p_1,p_2,0) + p_1^2
\Sigma^{(l)}_{2}(p_2,p_1,0)\Bigr) \ .
\end{eqnarray}
Thus, 
one has constructed the Green functions with insertions of $F \tilde
F$ (\ref{GaThetaAA}) uniquely from extension to local gauge coupling
independent from the existence of an anomalous Ward identity.

Now we turn to  the axial Ward identity.
For local gauge coupling the anomalous axial Ward identity takes the following
general form:
\begin{equation}
\Bigl( {\bf w}^{\rm axial} - \partial^\nu \df{V^\nu} \Bigr) \Ga = 
\sum_{l \geq 1} r^{(l)}  e^{2(l-1)}
[F^{\mn}(eA) \tilde F_{\mn}(eA)]\cdot \Ga + 
\mbox{ soft terms. } 
\end{equation}
Up to matter part contributions (\ref{Vdef})
the insertions on the right-hand-side are uniquely defined by gauge invariance.

When the coupling is local the first order is distinguished from
higher orders:  Only in this order the expression of the anomaly is a
total derivative. Hence, integrating the local Ward identity we find
\begin{equation}
{\cal W}^{\rm axial} \Ga = \sum_{l\geq 2} \intd 
r^{(l)}  e^{2(l-1)}
[F^{\mn}(eA) \tilde F_{\mn}(eA)]\cdot \Ga + \mbox{soft terms.}
\end{equation}
Differentiating this identity twice with respect to $A^\mu$ and
once  with respect to the local coupling, the construction yields
immediately  the
non-renormalization theorem of the Adler-Bardeen anomaly: Since the
right-hand-side is vanishing for a test with respect to photon fields 
\begin{equation}
\df{A^\mu(x_1)} \df {A^\nu (x_2)} \df{e(y)} {\cal W}^{\rm axial} \Ga
= 0
\end{equation}
we find  order by order
\begin{equation}
r^{(l)} = 0 \quad \mbox{for} \quad l\geq 2\ . 
\end{equation}

\section{The axial anomaly in non-abelian gauge theories}

The construction of the previous section can be immediately applied to non-abelian gauge theories.
We consider non-abelian gauge theories with a simple gauge group as
$SU(N)$ and left- and right-handed  fermionic matter fields
transforming under an
irreducible representation and its complex conjugate representation
respectively. Thus one has parity conservation and the theory is free
of gauge anomalies. 

Extending the gauge coupling to an external field and including a
space-time dependent $\Theta$ angle in the same way as
it was done for QED we find the following gauge invariant action:
\begin{eqnarray}
\label{GaYM}
\Ga_{\rm YM} & = & - \frac 1{4 }\Tr \intd   \Bigl( \frac 1{g^2}
G^{\mn}(gA) G_{\mn}(gA) +
 \Theta \epsilon^{\mn\rs} G_\mn (gA) G_\rs(gA) \Bigr)\nonumber \\
& & { }+ \intd \Bigl( ( i \psibar_L^T
 \bar \sigma ^\mu D_\mu \psi_L + i 
 \psi_R^T \sigma ^\mu D_\mu \psibar_R ) - m (\psi^T_L \psi_R + \psibar^T_L
\psibar_R) \Bigr)\ .
\end{eqnarray}
$A^\mu = A^\mu_a \tau _a$ are the gauge fields and
$\tau_a $ are the generators of the gauge group:
\begin{equation}
[\tau_a, \tau_b] = i f_{abc} \tau_c \ ,
\end{equation}
with the normalization 
\begin{equation}
\Tr (\tau_a \tau_b ) = \delta_{ab}\ .
\end{equation}
 The non-abelian field strength tensor and the covariant derivative
defined by
\begin{eqnarray}
G^\mn (A)
& = & \partial^\mu A^\nu - \partial^\nu A^\mu +i  [A^\mu, A^\nu] \ ,\nonumber \\
D^\mu \psi_L & = & (\partial^\mu  - g(x) i T_a A^\mu + i V^\mu)\psi _L
\ , \nonumber \\
D^\mu \psi_R & = & (\partial^\mu  - g(x) i T^*_a A^\mu + i V^\mu)\psi
_R \ .
\end{eqnarray}
In the definition of covariant derivatives 
we have already introduced the abelian external vector field $V^\mu$
whose
interaction with matter fields is
defined by the $U(1)$ axial Ward-identity (see (\ref{WIaxial}) and
(\ref{WOaxial})):
\begin{equation}
\Big({\mathbf w}^{\rm axial} - \partial^\nu \df{V^\nu}\Big) \Ga_{\rm YM} =
2 im  (\psi_L^T \psi_R - \psibar^T_L
\psibar_R) \ .
\end{equation}

 Quantization of non-abelian gauge theories cannot be performed by a
gauge Ward identity but one has to use  the corresponding BRS transformations.
For this purpose we rewrite the  gauge transformations into
BRS transformations by replacing the gauge transformation parameter
$\omega$ with the ghost field $c$
\begin{eqnarray}
\brs \phi &= & \delta^{\rm gauge}_c \phi, \qquad \phi\equiv A^\mu, \psi,
\psibar\ ,
\nonumber \\
\brs \Theta(x)&  = &
\brs g(x) = 
\brs V^\mu  = 0 \ ,
\end{eqnarray}
and introduce an anti-ghost and an auxiliary $B$-field with
\begin{equation}
\brs \cbar  = B\ ,  \qquad \brs B = 0\ .
\end{equation}
Then the gauge fixing term can be extended to a BRS-invariant action:
\begin{equation}
\Ga_{\rm g.f.} + \Ga_{\phi\pi} =
\brs \intd\Tr \bigl(\cbar \partial A\bigr)\ .
\end{equation}
Finally adding to the classical action the external field part,
\begin{equation}
\Gacl = \Ga_{\rm YM} + \Ga_{\rm g.f.} + \Ga_{\phi\pi} + \Ga_{\rm
ext.f.}\ ,
\end{equation}
with
\begin{equation}
\Ga_{\rm ext.f.} = \intd \Bigl(\rho^\mu \brs A_\mu + (Y_{\psi_L} \brs \psi_L
+ Y_{\psi_R} \brs \psi_R + \mbox{c.c.})\Bigr)
\end{equation}
we encode BRS transformations in the Slavnov--Taylor identity
\begin{equation}
{\cal S} (\Ga) = 0\ .
\end{equation}
The Slavnov--Taylor identity is valid to all orders also if we include
the local coupling and the $\Theta$ angle.

As for QED (\ref{Thetader})
the renormalization of the $\Theta$ angle is restricted  by
the equation 
\begin{equation}
\label{totalder}
\intd \df{\Theta } \Ga = 0\ ,
\end{equation}
which is a non-trivial restriction with local couplings.

 For the renormalization of the $\Theta$ angle in non-abelian
gauge  theories  we find the same results as for QED:
The only invariant counterterms to the $\Theta$ angle are the 
redefinitions of the axial vector field into the $\Theta$ angle (see
(\ref{GactTheta}) and
(\ref{Vdef})) and invariant counterterms to the $\Theta$ angle itself
are excluded by the identity (\ref{totalder})
(see (\ref{FtildeFct})).
 However, contrary to QED
the topological term itself
is renormalized by the 
wave-function renormalization of vector fields $A^\mu \to z_A A^\mu$.

Therefrom we conclude that
the insertions of the topological term into gluon self energies are
unambiguously determined by the symmetries of the theory, i.e.,
\begin{equation}
\label{brsGtildeG}
\brs_\Ga  \big[\Tr (G^\mn \tilde 
G_\mn) \big]\cdot \Ga = 0 \quad \mbox{and} \quad
\intd \big[ \Tr \, G^\mn \tilde G_\mn \big]\cdot \Ga = 0  \ .
\end{equation}
The explicit evaluation
 gives rise to  similar identities as (\ref{gaugeAA}) which involve in
addition the wave function renormalization of gluons. In order to get
 rid of the wave function renormalizations one can use  background
 gauge fields which satisfy a non-abelian gauge Ward
 identity. Insertions of the topological term into background self
 energies satisfy  the same identities of transversality
 as the photon fields (\ref{gaugeAA}).

With these ingredients the non-renormalization theorem of the axial
anomaly can be proven as  in QED. With local coupling the axial Ward
identity takes the general form:
\begin{equation}
\label{WIaxialna}
\Bigl( {\bf w}^{\rm axial} - \partial^\nu \df{V^\nu} \Bigr) \Ga = 
\sum_{l \geq 1} r^{(l)}  g^{2(l-1)}
[\Tr (G^{\mn}(gA) \tilde G_{\mn}(gA))]\cdot \Ga + 
\mbox{ soft terms } .
\end{equation}
All insertions on the right-hand-side are unambiguously defined by the
identities (\ref{brsGtildeG}). Integration of the local Ward identity
(\ref{WIaxialna}) 
yields then
\begin{equation}
r^{(l)} = 0 \quad\mbox{for} \quad l \geq 2\ ,
\end{equation}
i.e.\ the non-renormalization theorem of the Adler-Bardeen anomaly.

\section{Supersymmetric Yang-Mills theories}

Now we extend the Yang-Mills theories without matter to supersymmetric
theories: We introduce in addition to the gauge fields 
the
gaugino fields $\lambda^\alpha= \lambda_a^\alpha \tau_a$ and their
complex conjugate fields $\lambdabar^\alphadot$ and 
the auxiliary fields $D= D_a \tau_a$. For constant coupling the 
 action
\begin{equation}
\Ga_{\rm SYM} = \Tr \intd\Bigl(
- \frac 1{4g^2}  G^{\mn}(gA) G_{\mn}(gA) +  i 
 \lambda^ \alpha
 \sigma^\mu_{\alpha\alphadot} D_\mu \lambdabar ^ {\alphadot} +
\frac 1 8  
D^2 \Bigr)\ ,
\end{equation}
with
\begin{eqnarray}
D^\mu \lambda & = & \partial^\mu \lambda -  i g [A^\mu, \lambda]\ , 
\end{eqnarray}
is invariant under non-abelian gauge transformations and supersymmetry
transformations: 
\begin{equation} \begin{array}{rclcrcl}
\delta_{ \alpha}A^\mu &= &  i \sigma^\mu_{\alpha \alphadot}
\lambdabar ^{\alphadot }\ , & \qquad &\bar \delta_{ \alphadot}A^\mu &= & -
 i \lambda ^{\alpha }\sigma^\mu_{\alpha \alphadot}
\ ,   \\
\delta_{ \alpha} \lambda_{ \beta } & = &
- \frac i2 \big( \epsilon _{\alpha \beta} D 
- \sigma^{\mu \nu}_{\alpha \beta} G_{\mu \nu} \big)
\ ,& \qquad & \bar \delta_{ \alphadot} \lambda_\alpha & =& 0   \ ,\\
\delta_{ \alpha} \lambdabar_\alphadot &= & 0
\ ,&\qquad& \bar \delta_{ \alphadot} \lambda_{ \betadot } & = &
- \frac i2  \big(\epsilon _{\alphadot \betadot} D 
- \bar \sigma^{\mu \nu}_{\alphadot \betadot} G_{\mu \nu} \big)
\ ,  \\
\delta_{ \alpha}D & = & 2 \sigma^\mu_ {\alpha
\alphadot}
D_\mu \lambdabar ^ { \alphadot} \ ,& \qquad & 
\bar \delta_{ \alphadot}D & = & 2 
D_\mu \lambda ^ { \alpha}\sigma^\mu_ {\alpha
\alphadot} \ .
\end{array}\end{equation}

In the Wess-Zumino gauge \cite{WZ74_SQED,DWit75} that we have used here
the algebra of supersymmetry transformation does not close on translations but 
involves an additional field dependent gauge transformation:
\begin{eqnarray}
\{  \delta_{\alpha }, \bar \delta_{\alphadot} \}
&= &
2 i \sigma^{\mu}_{\alpha \alphadot} (\partial^\mu +
\delta^{\rm gauge}_{A^\mu}) \ ,\nonumber \\
\{  \delta_{\alpha },  \delta_{\beta} \} &= &
\{ \bar \delta_{\alphadot }, \bar \delta_{\betadot} \} = 0\ .
\end{eqnarray}
Thus, on gauge invariant field monomials the supersymmetry algebra
takes their usual form,  and it is  possible to classify the Lagrangians
into the usual $N=1$ multiplets.

The Lagrangians of supersymmetric field theories are the $F$ or $D$
components of a $N=1$ multiplet. In particular one finds that the
Lagrangian of the super-Yang-Mills action is the highest component of
a chiral and an antichiral multiplet. Using a superfield notation in the chiral
representation  the chiral Lagrangian multiplet $\L_{\rm SYM}$ is given by
\begin{eqnarray}
\label{LYMmult}
\L_{\rm SYM} 
& = &  - \frac 12 g^2  \Tr \lambda^ \alpha \lambda_\alpha + 
\Lambda^ \alpha  \theta_ \alpha+ \theta^ 2 L_{\rm SYM} \ ,
\end{eqnarray}
with the chiral super-Yang-Mills Lagrangian $L_{\rm SYM}$ 
\begin{eqnarray}
L_{\rm SYM} & = & \Tr \bigl(- \frac 14  G^{\mn}(gA) G_{\mn}(gA) +  i 
g \lambda^ \alpha
 \sigma^\mu_{\alpha\alphadot} D_\mu( g\lambdabar ^ {\alphadot}) + \frac 1 8  g^ 2
D^2 \nonumber \\
& & \quad
- \frac i 8 \epsilon^{\mu \nu \rho \sigma} G_{\mn}(gA) G_{\rho \sigma}
(gA) \bigr) 
\ . \label{LSYM}
\end{eqnarray}
By complex conjugation one obtains the 
   the respective antichiral multiplet $\bar {\cal L}_{\rm SYM}$.

The crucial point for the present considerations is the fact that the
topological term $\Tr\, G \tilde G$ appears in the supersymmetric
Lagrangians
 (\ref{LSYM})
and is
related to the kinetic term $- \frac 14 \Tr ( G G)$ via supersymmetry.
In previous sections we have shown that
 the higher order corrections to  $\Tr G \tilde G$  are unambiguously
determined  from gauge invariance, and therefore one expects the same
effect for the kinetic term, which describes the
renormalization of the gauge coupling. In ordinary perturbation theory these
relations are not found, since the topological term disappears from
the action by integration. However, with a local gauge coupling
one is able to include the complete Lagrangian 
with the topological term into the action and the
improved renormalization properties become apparent.

We will now extend the coupling constant $g$ to an external field
$g(x)$ \cite{KRST01,KR01}.
 For maintaining at the same time supersymmetry the coupling has
to be extended to a 
supermultiplet. Therefore we introduce a chiral and an antichiral
field multiplet 
$\etabold $ and $\etabarbold$:
\begin{equation}
\etabold= \eta + \theta \chi + \theta^2 f\ , \nonumber \\
\etabarbold= \etabar + \thetabar \chibar + \thetabar^2 \fbar \ ,
\end{equation}
which couple to the Lagrangian multiplet of the Super-Yang-Mills action:
\begin{eqnarray}
\Ga_{\rm SYM}
&= & -\frac 14
\intS \etabold \L_{\rm SYM} - \frac 14 \intSbar \etabarbold \bar \L_{\rm SYM}
\nonumber \\
& = & \intd \Bigl(\eta L_{\rm SYM} - \frac 12 \chi^\alpha \Lambda_\alpha - 
\frac 12 f g^2 \Tr \lambda^\alpha
\lambda_\alpha + \mbox{ c.c.}\Bigl)\ .
\end{eqnarray}
We identify the real part of $\eta $ with the inverse of the square of the
local  coupling:
\begin{equation}
\eta + \etabar = \frac 1 {g^2(x)}\ ,
\end{equation}
and observe that the imaginary part of $\eta$ couples to the
topological term.
  Thus, it takes the role of a space time dependent
$\Theta$ angle 
\begin{equation}
\eta - \etabar = 2 i \Theta\ .
\end{equation}
Hence, dependence on the superfield $\etabold$ and $\etabarbold$ is
governed by a topological formula (cf.~(\ref{topfor})) and  by the
identity
(\ref{totalder}): 
\begin{equation}
\label{holomorph}
\intd \Big(\df{\eta} - \df{\etabar}\Big) \Ga
= -i \intd \df{\Theta} \Ga = 0\ .
\end{equation}

Constructing with these ingredients the invariant 
counterterms to $\Ga_{\rm SYM}$ we find
from gauge invariance and
supersymmetry:
\begin{eqnarray}
\label{GaSYMct}
\Ga^{(l)}_{\rm ct, phys} 
&= & z^{(l)}_ {\rm YM}
\Bigl(-\frac 18
\intS \etabold^{1-l} \L_{\rm SYM} - \frac 18 \intSbar
\etabarbold^{1-l} \bar \L_{\rm SYM} \Bigr)
\nonumber \\
& = & z^{(l)}_{\rm YM}
\intd \Bigl( - \frac 14 (2g^2)^{l-1} G^\mn(gA) G_\mn(gA)  \nonumber
\\
& & \phantom{z^{(l)}_{\rm YM}
\intd \Bigl( -}
-\frac 18 (2g^2)^l (l-1) \Theta G^\mn(gA) \tilde G_\mn(gA) +
\ldots
\Bigr)\ .
\end{eqnarray}
These counterterms as well as the respective UV divergences are excluded
in
loop order $l \geq 2$ by the identity (\ref{holomorph}), which governs
the renormalization of the $\Theta$ angle:
\begin{equation}
\label{ctholomorph}
z^{(l)}_{\rm YM} = 0 \qquad \mbox{for} \quad l \geq 2 \ .
\end{equation}
Thus, in loop orders $l\geq 2$ the topological term
$\Tr\,  G\tilde G$ determines the
renormalization of the coupling.
It is obvious from eq.~(\ref{GaSYMct}) that the  one-loop order is
special, and
  we will show in the next section that there
 the loop corrections to $\Tr\, G \tilde G$
induce an anomaly of  supersymmetry.

\section{The anomalous breaking of supersymmetry}

To quantize supersymmetric field theories in the Wess-Zumino gauge
one includes the gauge transformations, supersymmetry
transformations and translations into the nilpotent BRS operator
\cite{White92a,MPW96a}:
\begin{equation}
\brs \phi = \delta^{\rm gauge}_c + \epsilon^\alpha \delta_\alpha +
\bar \delta_\alphadot \epsilon^\alphadot - i  \omega^\nu \delta_\nu^T\ .
\end{equation}
The fields $c(x)$ are the usual Faddeev-Popov ghosts, $\epsilon^\alpha$,
$\epsilonbar^\alphadot$ and $\omega^\nu$  are
constant ghosts of supersymmetry and translations, respectively.

As for usual gauge theories BRS transformations are encoded in the
Slavnov-Taylor identity:
\begin{equation}
\label{ST}
{\cal S}(\Gacl) = 0
\end{equation}
with
\begin{equation}
\Gacl = \Ga_{\rm SYM} + \Ga_{g.f.} + \Ga_{\phi\pi} + \Ga_{ext.f.}\ .
\end{equation}
Details of the construction are not relevant for the following
considerations, but it is only important to keep in mind that the
Slavnov--Taylor identity for supersymmetric actions
expresses gauge invariance as well as
supersymmetry.

In addition to the Slavnov-Taylor identity the dependence on the
external fields $\etabold$ and $\etabarbold$ is restricted 
to all orders by the
identity (\ref{holomorph}) \cite{KR01}.

In the course of renormalization 
the Slavnov-Taylor identity (\ref{ST}) has to be established
 for the 1PI Green functions to all
orders of perturbation theory.
From the quantum action principle one finds that the possible breaking
terms
are local in one-loop order:
\begin{equation}
{\cal S}(\Ga) = \Delta_{\rm brs} + {\cal O}(\hbar^2)\ .
\end{equation}
Algebraic consistency yields  the  constraint
\begin{equation}
\brs_{\Gacl} \Delta_{\rm brs} = 0 \ .
\end{equation}
Gauge invariance can be established as usually, i.e.\ one has
\begin{equation}
{\cal S}(\Ga)\Big|_{\epsilon,\epsilonbar = 0}
 =  {\cal O}(\hbar^2) \ ,
\end{equation}
and  the remaining breaking terms depend on the supersymmetry ghosts
$\epsilon$ and $\epsilonbar$, and represent as such a breaking of
supersymmetry. 
Having gauge invariance established, the supersymmetry algebra closes
on translations. Using the supersymmetry algebra one obtains that the
remaining  breaking terms of supersymmetry are
variations of field monomials with the quantum numbers of the action:
\begin{equation}
\Delta_{\rm brs} = \brs_{\Gacl} \hat \Ga_{\rm ct}\ .
\end{equation}
However, not all of the field monomials in $\hat \Ga_{\rm ct} $
represent scheme-dependent counterterms of the usual form.
 There is one  field monomial in $\hat \Ga_{\rm ct} $, which
 depends on the logarithm of the gauge coupling, but whose BRS variation
 is free of logarithms:
\begin{eqnarray}
\label{Deltaanomalybrs}
\Delta^{\rm anomaly}_{\rm brs} & =   & \brs \intd \ln g(x) (L_{\rm SYM} + \bar
L_{\rm SYM}) \\
& = & (\epsilon^ \alpha \delta_\alpha + 
\epsilonbar^ \alphadot \bar \delta_\alphadot) 
\intd \ln g(x) (L_{\rm SYM} + \bar
L_{\rm SYM}) \nonumber\\
& = & \intd  \Bigl(i\; \ln g(x)  \bigl(\partial _\mu \Lambda ^
\alpha \sigma ^ \mu_{\alpha \alphadot} \epsilonbar^ \alphadot -
\epsilon^ \alpha \sigma^ \mu_{\alpha \alphadot}  \partial_\mu \bar \Lambda
^ {\alphadot} \bigr) \nonumber \\
& & \phantom{\intd} - \frac 12 g^ 2 (x) (\epsilon \chi + \chibar
\epsilonbar)
(L_{\rm SYM} + \bar L _{\rm SYM}) \Bigr)\nonumber\ .
\end{eqnarray}
Indeed, due to the total derivative in the first line the breaking
$\Delta^{\rm anomaly}_{\rm brs}$ is free of logarithms 
for constant coupling and for any test with respect to the local
coupling. Moreover one immediately verifies, 
that  $\Delta^{\rm anomaly}_{\rm brs}$ satisfies
the topogical formula in one-loop order. 
Therefore 
$\Delta^{\rm anomaly}_{\rm brs}$  satisfies all constraints on the
breakings 
and can appear
as a breaking of the 
 Slavnov-Taylor identity in the first order of perturbation theory.

However, being the variation of a field monomial depending on the
logarithm of the coupling
 $\Delta^{\rm anomaly}_{\rm brs}$ cannot be induced by divergent
one-loop diagrams, which are all power series in the coupling.
Thus,
the corresponding
counterterm is not related to a naive contribution induced in the
procedure of subtraction and does not represent  a naive redefinition of
time-ordered 
Green functions. 
 Therefore,
 $\Delta^{\rm anomaly}_{\rm brs}$ is an anomalous breaking of supersymmetry 
 in perturbation theory and we remain with
\begin{equation}
{\cal S}(\Ga) = r_\eta^{(1)} \Delta^{\rm anomaly}_{\rm brs} + {\cal
O}(\hbar^2)\ .
\end{equation}

From its characterization it is straightforward to prove with algebraic methods
 that
the coefficient of the anomaly  is gauge and
scheme independent \cite{KR01}. 
Furthermore,
by evaluating the Slavnov--Taylor identity one can find an expression
for $r_\eta^{(1)}$ in terms of convergent loop integrals \cite{KR01anom}. 
 Using 
background gauge fields $\hat A ^\mu$ and Feynman gauge $\xi = 1$ 
 the anomaly coefficient is
explicitly related to insertions of the topological term and
 the axial current of gluinos into
  self energies of background fields:
\begin{equation}
\label{result}
 g^2 r^ {(1)}_\eta = - \frac 12 \Sigma^ {(1)}_{\eta-\etabar}(p_1,-p_1) \Big|_{ \xi=1} \ ,
\end{equation}
where $\Sigma_{\eta -\etabar}$ is defined by
\begin{eqnarray}
\Ga_{\eta - \etabar \hat A^\mu_a \hat A_b^\nu}(q, p_1,p_2)
&= &  \Bigl(\bigl[i\; \Tr \bigl(\partial( g^2 \lambda \sigma \lambdabar) - \frac 14 G^
{\mu\nu}\tilde G_{\mu \nu}(gA +\hat A) \bigr)\bigr] \cdot \Ga
\Bigr)_{\hat A_a^\mu \hat A_b^\nu}
(q,p_1,p_2)\nonumber \\ 
&   & { } = \; i \epsilon^ {\mu \nu \rho \sigma}p_{1 \rho } p_{2 \sigma} \delta_{ab}
\bigl(-2 + \Sigma_{\eta
- \etabar} (p_1,p_2)\bigr) \ .
\end{eqnarray}
In section 2 and 3 these Green functions have been constructed 
 in abelian as well as non-abelian gauge theories by
 extension to local gauge coupling in terms of convergent integrals
(see (\ref{GaThetaresult}) and (\ref{brsGtildeG})).
Due to the Ward identity for the background gauge fields,
\begin{equation}
\bigl({\mathbf w}^{\rm gauge}_a - \partial^\nu \df{\hat A^\nu_a}\bigr) \Ga = 0
\end{equation}
the 1PI Green functions $\Ga_{\eta-
\etabar\hat A^\mu_a \hat A^\nu _b}$
 can be even constructed 
 in the same
way as it was demonstrated for $\Ga_{\Theta A^\mu A^\nu}$ in QED (\ref{GaThetaresult}).

Explicit evaluation  of the respective one-loop diagrams yields
\begin{equation}
\label{value}
r^ {(1)}_\eta = (-1 + 2)\frac {C(G)}   {8 \pi^2} =\frac {C(G)}   {8 \pi^2} \ ,
\end{equation}
where the first term comes from the axial anomaly of gauginos and
the second term from the insertion of the topological term.

We want to mention that the anomaly coefficient vanishes in SQED,
since one-loop diagrams to $\Ga_{\eta- \etabar  A^\mu  A^\nu}$ do not
exist. The anomaly coefficient also vanishes in $N= 2$ theories ,
since there the two  fermionic fields just cancel the contribution arising
 from the topological term $\Tr\, G^\mn \tilde G_\mn$.

\section{Implications of the anomaly}

In the framework of perturbation theory
the anomaly of supersymmetry cannot be removed by a local
counterterm. However, one is able to proceed with algebraic
renormalization nevertheless by rewriting the anomalous breaking in
the form of a differential operator \cite{KR01}:
\begin{eqnarray}
\label{STanom1loop}
\Delta^{\rm anomaly}_{\rm brs}& = &
\intd \Bigl(g^ 6 r_\eta^{(1)}(\epsilon
\chi + \chibar \epsilonbar) \df {g^2} \nonumber  
\\ 
& & \qquad { }- 
i  r_\eta^{(1)} 
 \partial_\mu \ln g^ 2 \bigl( (\sigma ^ \mu \epsilonbar)^\alpha
\df {\chi^ \alpha}  + (\epsilon \sigma^ \mu )^ \alphadot \df {\chibar^
 \alphadot} \bigr)\Bigr) \Gacl
\end{eqnarray} 
Then one has
\begin{equation}
\big({\cal S} + r^{(1)}_\eta \delta {\cal S}\big) \Ga = {\cal
O}(\hbar^2)\ .
\end{equation}
For algebraic consistency one has to require the nilpotency properties
of the classical Slavnov-Taylor operator also for the extended
operator. Nilpotency  determines an algebraic consistent
continuation of (\ref{STanom1loop}). This continuation is not unique
but contains at the same time all redefinitions of the coupling
compatible with the formal power series expansion of perturbation
theory. 

As a result one finds an algebraic consistent
Slavnov-Taylor identity  in presence of the anomaly:
\begin{equation}
\label{STreta}
{\cal S}^ {r_\eta} (\Ga) = 0 \quad \mbox{and} \quad \intd \Bigl(\df{\eta} -
\df{\etabar} \Bigr)\Ga = 0\ ,
\end{equation}
where the anomalous part is of the form:
\begin{eqnarray}
{\cal S}^ {r_\eta} (\Ga) = {\cal S}(\Ga) & - & \intd \Bigl(g^ 4 \delta F(g^ 2)(\epsilon
\chi + \chibar \epsilonbar) \df {g^2} \nonumber \\ & &{ } \quad + 
i \frac {\delta F }{1+ \delta F}
 \partial_\mu g^ {-2}  \bigl( (\sigma ^ \mu \epsilonbar)^\alpha
\df {\chi^ \alpha}  + (\epsilon \sigma^ \mu )^ \alphadot \df {\chibar^
 \alphadot} \bigr)\Bigr)\Ga \ ,
\end{eqnarray}
with
\begin{equation}
\delta F(g^2)  = r^{(1)}_\eta g^2  + {\cal O}(g^4) \ .
\end{equation}
The lowest order term is uniquely fixed by the anomaly, whereas
the higher orders in $\delta F(g^2) $  correspond to the
scheme-dependent finite redefinitions of the coupling.

The simplest choice for $\delta F$ is given by
\begin{equation}
\label{Fmin}
 \delta F =  r_{\eta}^{(1)} g^2\ ,
\end{equation}
and another  choice is provided by
\begin{equation}
\label{FNSZV}
\frac{\delta F}{1 + \delta F} =  r_{\eta}^{(1)} g^2\ .
\end{equation}
As seen below, the latter choice gives the NSZV expression of the
gauge $\beta$ function \cite{NSVZ83,SHVA86}.

Algebraic renormalization with the symmetry operator (\ref{STreta}) is
performed in the conventional way.
In particular one can derive the $\beta$ functions from an algebraic
construction of the renormalization group equation in presence of
the local coupling. 
Starting from the classical expression of the RG equation
\begin{equation}
\kappa\partial _\kappa  \Gacl = 0
\end{equation}
we construct the higher orders of the RG equation
by constructing the  general basis of
symmetric differential operator with the quantum numbers of the action:
\begin{equation}
{\cal R} = \kappa \partial _\kappa + {\cal O}(\hbar) \ ,
\end{equation}
with
\begin{eqnarray}
& & \brs _\Ga^{r_\eta}{\cal R} \Ga -   {\cal R} {\cal S}^{r_\eta}(\Ga) =
 0
\ ,\label{RST} \\
& & \Big[\intd \Bigl(\df{\eta} - \df \etabar\Bigr), {\cal R} \Big] =
0\ .\label{Rholomorph}
\end{eqnarray}
The general basis for the symmetric differential operators
consists of the differential operator of the supercoupling
$\etabold$ and $\etabarbold$ and several field redefinition operators.
The differential operator of the coupling determines the $\beta$
function of the coupling, whereas the field redefinition operators
correspond to the anomalous dimensions of fields.

For the present paper
we focus on the operator of the $\beta$ function and neglect the
anomalous dimensions. For proceeding
we construct first the RG operator, which is symmetric with respect to the classical
Slavnov-Taylor operator, i.e.\  we set $\delta F = 0$.
Using  a superspace  notation we find:
\begin{equation}
{\cal R}_{\rm cl} = -
\sum_{l \geq 1} \hat \beta_g^{(l)} 
\Bigl(\intS 
\etabold^{-l+1}\df {\etabold } + \intSbar \etabarbold^{-l+1}\df
{\etabarbold}
\Bigr) + \ldots\  ,
\end{equation}
and
\begin{eqnarray}
& & \brs _\Ga{\cal R}_{\rm cl} \Ga -   {\cal R}_{\rm cl} {\cal S}(\Ga) =
 0\ .
\end{eqnarray}
Evaluating then the consistency equation~(\ref{Rholomorph}) we obtain 
\begin{equation}
\hat\beta_g^{(l)} = 0  \quad \mbox{for } \quad l \geq 2 \ .
\end{equation}
These restrictions are the same restrictions as we have found for the
invariant counterterms of the super-Yang-Mills action (see 
(\ref{GaSYMct} with)
(\ref{ctholomorph})) and represent
the fact, that for super-Yang-Mills the renormalization of the
coupling is governed by the 
renormalization of the $\Theta$ angle in loop orders greater than one.

Finally we extend the one-loop operator ${\cal R}_{\rm ct}$
to a symmetric operator with
respect to the anomalous ST identity (\ref{RST}) and find by a straightforward
calculation 
\begin{eqnarray}
\label{Rgen}
{\cal R} & =&  \hat \beta_g^{(1)} \intd g^3 (1 + \delta F(g^2)
)
\df{g} + \ldots \nonumber \\
& = &  \hat \beta_g^{(1)} \intd g^3 \big(1 + r_\eta^{(1)}g^2 + {\cal
O}(\hbar^2)\big) 
\df{g} + \ldots \ .
\end{eqnarray}
For constant coupling we find from (\ref{Rgen}) the closed expression
of the gauge $\beta$ function
\begin{equation}
\beta_g = \hat \beta_g^{(1)}  g^3 (1 + \delta F(g^2) ) =
\hat \beta_g^{(1)}  g^3 \big(1 + r_\eta^{(1)}g^2 + {\cal
O}(\hbar^2)\big) \ .
\end{equation}
Thus, the two-loop order is uniquely determined by the anomaly and the
one-loop coefficient, whereas higher orders depend on the specific
form one has chosen for the function $\delta F(g^2)$. In particular
one has for the minimal choice (\ref{Fmin}) a pure two-loop
$\beta$-function and for the NSZV-choice (\ref{FNSZV}) one obtains the NSZV
expression
\cite{NSVZ83} of the gauge $\beta$ function of pure Super-Yang-Mills
theories. 

\section{Conclusions}

The extension of the gauge coupling to an external field has been
shown to be a crucial step for determining the renormalization of the
topological term. As its definition is a necessary prerequisite
 for the proof of the Adler-Bardeen
non-renormalization  its quantum corrections have been considered in the
past with
different methods \cite{ZA90,Reuter}. In a strict sense, however,
these corrections
are determined in
perturbation theory with
constant coupling only in combination with an anomalous chiral
symmetry \cite{BMS84,PISI87AB, BOS92}.
 
With local coupling, however, the situation is drastically changed:
Using gauge invariance with local coupling  it turns
out that  
renormalization of the topological term
is uniquely fixed by convergent one-loop integrals in
the same way as it is for the triangle diagrams which induce the
Adler-Bardeen anomaly. 

In the present paper we have given two implications of the construction:
First, a proof of the non-renormalization of the Adler--Bardeen
anomaly, and  second, we have shown
 that the quantum corrections to the topological term induce
an anomaly of supersymmetry in
supersymmetric Yang-Mills theories 
 in one loop order and determine the renormalization of the coupling
in all  loop orders greater than one.        

\vspace{0.5cm}
{\bf Acknowledgments}

I wish to thank the organizers of the workshop, F. Scheck,
R. Nest and E. Vogt, for the invitation and for the organization of
this stimulating meeting between mathematicians and physicists.


\end{document}